# Quantifying the effect of image compression on supervised learning applications in optical microscopy

Enrico Pomarico[1*], Cédric Schmidt[1], Florian Chays[1], David Nguyen[2], Arielle Planchette[2], Audrey Tissot[4], Adrien Roux[1], Stéphane Pagès[4,5], Laura Batti[4], Christoph Clausen[6], Theo Lasser[3], Aleksandra Radenovic[2], Bruno Sanguinetti[6], and Jérôme Extermann[1*]

[1]HEPIA/HES-SO, University of Applied Sciences of Western Switzerland, Rue de la Prairie 4, 1202 Geneva, Switzerland

[2]Laboratoire de Biologie à l'Echelle Nanométrique, School of Engineering, École Polytechnique Fédérale de Lausanne, CH-1015 Lausanne, Switzerland

[3]Max-Planck Institute for Polymer Research, Ackermannweg 10, 55128 Mainz, Germany

[4]Wyss Center for Bio- and Neuroengineering, Geneva, Switzerland

[5]Department of Basic Neurosciences, Geneva Neuroscience Center, Faculty of Medicine, University of Geneva, Geneva, Switzerland.

[6]Dotphoton SA, Zeughausgasse 17, 6300 Zug, Switzerland

[*]email: enrico.pomarico@hesge.ch; jerome.extermann@hesge.ch

**The impressive growth of data throughput in optical microscopy has triggered a widespread use of supervised learning (SL) models running on compressed image datasets for efficient automated analysis. However, since lossy image compression risks to produce unpredictable artifacts, quantifying the effect of data compression on SL applications is of pivotal importance to assess their reliability, especially for clinical use.**

**We propose an experimental method to evaluate the tolerability of image compression distortions in 2D and 3D cell segmentation SL tasks: predictions on compressed data are compared to the raw predictive uncertainty, which is numerically estimated from the raw noise statistics measured through sensor calibration.**

**We show that predictions on object- and image-specific segmentation parameters can be altered by up to 15% and more than 10 standard deviations after 16-to-8 bits downsampling or JPEG compression. In contrast, a recently developed lossless compression algorithm provides a prediction spread which is statistically equivalent to that stemming from raw noise, while providing a compression ratio of up to 10:1. By setting a lower bound to the SL predictive uncertainty, our technique can be generalized to validate a variety of data analysis pipelines in SL-assisted fields.**

In the last years, an ever-growing community of optical microscopists is facing a massive data throughput, long-term storage costs, data transfer limitations and, more importantly, the need for automated data analysis, which has paved the way for extensive use of artificial intelligence (AI) methods. Supervised learning (SL) algorithms are routinely adopted to automate classification, segmentation, and artificial labelling of cellular or sub-cellular structures[1,2,3], biological tissues[4,5,6], as well as material defects[7,8,9]. These SL approaches have reported remarkable results in various fields, such as medical screening[10,11], single molecule localization[12,13] and drug discovery[14,15]. Deep-learning (DL) algorithms have also been successfully employed for micrograph restoration, in particular for de-noising and spatial resolution enhancement[16,17,18].

However, to deal with large training datasets and computational power constraints, SL models are ubiquitously executed on compressed imaging datasets. Despite the fact that they produce visually faithful images, lossy compression algorithms introduce artifacts and remove an unpredictable amount of information from the original raw image, leading image processing and analysis, often performed several times after archiving, to unreliable outcomes. Whether image quality degradation induced by lossy compression has detrimental effects on the results of SL tasks is an open question, and experimental methods are needed to quantify the effect of compression distortions on predictions and performances of SL applications.

The difficulty of assessing the impact of a processing pipeline on the outcomes of SL applications is related to the more general problem of quantifying their predictive uncertainty, which constitutes a necessary step to enable AI-assisted decision-making, especially for clinical use[19]. The lifecycle of an

AI process is affected by multiple sources of model- and data-driven uncertainties[19]. Some theoretical Bayesian deep-learning models have been introduced to provide probabilistic interpretations to model outputs[20,21]. However, these approaches focus on model uncertainties and ask for high computational costs, while providing lower prediction performances compared to standard DL approaches[21]. On the other hand, data noise seems to affect predictive variance more significantly than model-driven uncertainties[20].

In this regard, raw data, typically obtained through a digitization operation on a physical sample via an acquisition instrument, followed by a few minimal corrections, are intrinsically affected by the noise associated with the acquisition process. In optical microscopy, raw data variability is mainly provided by the quantum noise of the photons hitting the sensor, as well as by the electronic noise[22]. Hence, if one performs a sample acquisition under stabilized illumination conditions, as shown in Figure 1, the acquired raw images are not identical and the raw pixel values display a statistical distribution of average $\mu$ and width $\sigma$ (the standard deviation associated to the per-pixel noise).

In this work, we propose an experimental method to decipher the predictive uncertainty in SL-based optical microscopy applications, relying on the statistical noise of raw imaging data. Raw statistics could be in principle determined by repeating and averaging the acquisition of the same image a number of times. However, these tests are often hard to be carried out in a microscopy laboratory, because of sample variability, non-stationary illumination conditions, as well as computational limitations. As a first step, we implement a calibration of the microscope's imaging sensor to determine an accurate mapping between light intensities and per-pixel noise. This is performed by sending to the sensor variable intensities from a homogeneous and controlled light source (see Supplementary Material). Secondly, using the sensor-dependent calibration data, statistically raw-equivalent images are generated by replacing the original pixel value with a random one satisfying the raw pixel value statistical distribution. By testing a previously trained SL model on the synthetic images, the prediction spread $\sigma_{raw}$ associated to a certain parameter $\chi$ can be determined (Figure 1), similarly to a Monte-Carlo simulation.

Estimating the confidence level of SL predictions constitutes a strategy to validate any kind of processing performed on images used to make predictions. In particular, we will use our method to evaluate whether image compression leads to predictions exceeding the original predictive uncertainty $\sigma_{raw}$. To this end, the last step of our method consists in comparing $\sigma_{raw}$ to the difference between the values of a parameter χ without ($\chi_{raw}$) and with compression ($\chi_c$), by using, as a figure of merit, the standard score defined as

$$\epsilon = \frac{\chi_{raw} - \chi_c}{\sigma_{raw}}.$$

A standard score $\epsilon$ such that $|\epsilon| < 1$ indicates that the statistical variability of SL predictions is lower than that stemming from the natural noise level of raw micrographs, thus defining a general criterion of tolerability of image compression distortions.

As noise in raw data is unavoidable, this approach allows one to estimate the minimal level of uncertainty in SL-assisted decision making. In principle, other sources of variability, such as representational, labeling, as well as model uncertainties[19], should be added to estimate the uncertainty of a complete AI-decision making pipeline. In addition, our experimental method can be generalized to any image processing operation, as well as any field where acquisition sensors can be calibrated and data undergo pre-processing before AI use.

To demonstrate our method, we perform SL-based cellular segmentation tasks on 2-dimensional (2D) image datasets, obtained through phase-contrast (PC) microscopy, as well as 3-dimensional (3D) datasets obtained via light-sheet (LS) microscopy and optical projection tomography (OPT). By increasing the dimensionality of the raw dataset, we aim at observing how the complexification of raw data noise impacts the SL predictive uncertainty.

We aim at evaluating digital distortions induced by three types of operations aimed at reducing data volume: 16-to-8 bit conversion, 10:1 JPEG compression, as well as an efficient lossless compression technique developed by the Dotphoton (DP) company[23], suited to computer vision-based tasks in microscopy. The DP method reports compression ratios from 5:1 to 10:1 after an initial image preparation step, in which image noise is replaced with a pseudo-random noise that closely mimics the statistical distribution of the raw pixel values. Although the noise replacement reduces the signal-to-noise ratio of each pixel by 1.2 dB, it allows to achieve high compression factors, since the pseudo-noise can be computed and makes the subsequent application of a standard lossless compression algorithm more efficient[24].

We first consider an SL-based cell segmentation application performed on images of microspheres, as well as mouse kidney collecting duct (MPK) cells, acquired with a PC microscope. A random forest (RF) algorithm[25], taking decisions on the basis of morphological spatial features, is trained via a manual pixel-based labelling performed on a raw image. After producing a segmentation map, we estimate parameters related to the whole image, as well as specific to single segmented objects. After calibration of the acquisition camera, we generate a set of 10 statistically raw-equivalent images and determine the predictive uncertainty $\sigma_{raw}$ of the considered parameters as the standard deviation of the values obtained with this dataset. The segmented mask obtained from an image of microspheres is shown in Figure 2a. In Figure 2d, we compare the number of counted objects $N_{tot}$ and the total segmented area $A_{tot}$ predicted from the raw image with those obtained from the corresponding 7:1 DP, 8-bit and 10:1 JPEG file. The difference in $N_{tot}$ with respect to the raw value is in all cases between 0.2% and 0.5% ($\epsilon \approx 1$ in all cases). In contrast, the predicted value of $A_{tot}$ in the 8-bit and JPEG case ($\epsilon = 62$ in both cases) shows a 2% deviation with respect to the raw result (10 times larger than the spread caused by the raw noise (0.2%)), while only 0.1% variation in the DP case ($\epsilon = -2$).

We then perform a similar analysis on 19 different morphological parameters estimated for each segmented object (area, center of mass coordinates ($X_{CM}$, $Y_{CM}$), perimeter, major axis, minor axis, ellipsoid angle, circularity, Feret, Feret X, Feret Y, Feret angle, Minimum Feret, aspect ratio, roundness, solidity, Feret aspect ratio, compactness, and extent). The inset of Figure 2b shows a linear trend between the single-object area estimated from the raw ($A_{raw}$) and the DP ($A_{DP}$) mask. The points accumulating along the diagonal correspond to aggregates of more than one microsphere. The distribution of the difference in the single-object area obtained from the synthetic raw-equivalent images ($\Delta A = A_{raw} - A_{synt\,raw}$) (mean: -0.01 pixels$^2$, standard deviation: 1.33 pixels$^2$) is in good agreement with the one obtained from the DP format ($\Delta A = A_{raw} - A_{DP}$) (mean: -0.05 pixels$^2$, standard deviation: 1.01 pixels$^2$). However, a clear shift of the distribution of $\Delta A$ is observed in the 8-bit and JPEG case (mean: 2.9 pixels$^2$, standard deviation: 3.4 pixels$^2$ for 8-bit and respectively mean: 3.1 pixels$^2$, standard deviation: 3.5 pixels$^2$ for JPEG) (Figure 2c). Notice that the spread of the synthetic raw distribution shown in Figure 2c, corresponding to the predictive uncertainty due to the intrinsic noise of the raw image, is around 1 pixels$^2$, which is smaller than that provided by the Point Spread Function (PSF) of the microscope (see Supplementary Material). In Figure 2e, we plot the standard scores for all parameters averaged over all objects. All scores for the DP format are close to zero and have standard deviations (indicated by error bars) of the order of 1. This result indicates that alterations induced by the DP compression can be considered statistically equivalent to those produced by the intrinsic noise of the raw images. In contrast, the distribution of the standard scores in the 8-bit and JPEG case is larger than the [-1,1] interval, with average often far from 0 for almost all parameters. This implies that predictions on the 8-bit and JPEG file exceed the predictive uncertainty provided by the raw statistical noise. As expected, we observe that JPEG files with stronger compression deviate dramatically from the raw predictions. By reducing the compression factor, these deviations decrease and get progressively close to those provided by the 8-bit file.

The same analysis is performed on a PC micrograph of MPK cells with high confluence and high mean pixel intensity. In these experimental conditions, the previously reported good agreement between the DP and the raw results is confirmed, and we once again observe a 1-2% deviation in the values of $A_{tot}$ in the 8-bit and JPEG cases (Figure 2g). Results from the single object parameters (Figure 2h) show that the statistical equivalence of processing results on raw and DP data does not depend on the spatial configuration of the segmented objects. Additionally, the single-object parameters turn out to be significantly altered in the 8-bit and JPEG case.

We then challenge our method on more complex segmentation tasks involving 3D imaging datasets produced by LS microscopy and OPT. Via the estimation of the raw noise-related AI predictive uncertainty, we study the impact of image compression on these applications. This is particularly relevant to justify the use of compression for high volume data management issues and computer-vision automated tasks.

By the use of a RF voxel classification algorithm, we perform segmentation of neuronal nuclei in a portion of a mouse brain after c-*fos* antibody staining, imaged by a mesoscale Selective Plane Illumination LS Microscope (mesoSPIM) with 1x objective (Figure 3a and b)[26] and providing a 4.3 GB raw dataset. To obtain the 3D segmented volumes, a threshold criterion is applied to the 3D probability map obtained from the trained ML model. We perform these tests only with DP compression, providing a 7.3:1 compression ratio, and 8-bit conversion.

As shown in Figure 3f, the number of segmented nuclei $N_{tot}$ (around 190'000), as well as the total segmented volume $V_{tot}$ and surface area $SA_{tot}$ calculated from the DP dataset, display a discrepancy of 0.1-0.2% with respect to the values obtained from the raw segmented 3D image ($\epsilon = -1.6, -0.9, -0.1$ for $N_{tot}$, $V_{tot}$ and $SA_{tot}$ respectively) and are perfectly compatible with the spread shown by the synthetic images. In contrast, for all global parameters, the 8-bit conversion provides 15-20% discrepancy, much larger than in the 2D case ($\epsilon = 366, 87, 9$ for $N_{tot}$, $V_{tot}$ and $SA_{tot}$ respectively).

The 2D histogram in Figure 3d compares the volume in voxels (1 voxel = 5.26 μm x 5.26 μm x 5 μm = 138 μm$^3$) of each single object identified in the raw ($V_{raw}$) and in the DP dataset ($V_{DP}$). To identify the same corresponding objects in the two datasets, especially those that do not have the same center coordinates after segmentation, an association based on the minimal Euclidean distance is performed. The concentration of the points along the diagonal is an indication of good prediction of the single-object volume from the segmented DP dataset, in contrast to the same distribution plotted for the 8-bit case (Figure 3e). The distribution of the single-volume differences $\Delta V$, displayed in Figure 3g, obtained from the synthetic raw files ($\Delta V = V_{raw} - V_{synt\ raw}$) (mean: 0.06, standard deviation: 2.7) is in good agreement with that provided by the DP stack ($\Delta V = V_{raw} - V_{DP}$) (mean: -0.06, standard deviation: 2.2). Also, in the 3D case the raw spread turns out to be smaller than the PSF of the LS microscope (see Supplementary Material). Interestingly, the 8-bit conversion alters the distribution of volume differences in a more dramatic way compared to the 2D case (mean: 7.3, standard deviation: 6.4) (Figure 3e), confirming the results obtained for the global parameters. The 8-bit conversion seems to affect the action of the morphological operators used in the SL algorithm in a more complex way with respect to the 2D case. This effect is probably due to the higher noise correlation existing between the different slices of the 3D LS microscopy dataset.

We then adopt our compression tolerability method to determine the uncertainty of SL outcomes in a pre-clinical application, based on the estimation via Optical Projection Tomography (OPT) imaging of amyloidosis in a mouse brain affected by Alzheimer's disease. OPT is well suited to image mesoscopic centimeter-sized biological specimens, such as organs, and represents the optical equivalent of computed tomography: fluorescent projection images are captured at different angles around the specimen and the 3D image of the organ is reconstructed via a Filtered Back Projection (FBP) algorithm using an inverse Radon transform (Figure 4a, b, c).[27] Compared to the segmentation performed on LS microscopy data, in OPT an image reconstruction step is introduced before the AI testing, rendering the propagation of the original noise of the raw projections through the AI pipeline more complex. To obtain the amyloid plaques segmentation mask on the raw and compressed 3D images, we adopt the same RF algorithm used for the LS microscopy data to classify every voxel of the 3D reconstructed images and apply a threshold criterion on the 3D probability map (namely 0.7 for brain anatomy and 0.5 for amyloid plaques) according to a previous study[28]. A close up of a slice of a reconstructed diseased mouse brain and the corresponding plaques segmentation mask are shown in Figure 4d and e, respectively. In this case, to quantify the SL predictive uncertainty, we simulate a dataset of 10 statistically raw-equivalent projections of a single middle-aged mouse brain.

The voxel classifier relies on intensity-based, edge-based, as well as texture operators, computed on the 3D image after various levels of Gaussian smoothing. We first compare the results of all operators on the raw dataset of 0.5 GB size, the 7.8:1 DP compressed and the 8-bit converted projections datasets, by calculating their standard scores averaged over all projection pixels. As shown by Figure 4f, the averaged standard scores belong for all parameters to the [-1,1] interval in the DP case. In contrast, the 8-bit conversion shows in general larger discrepancies for an increasing gaussian smoothing. A similar result can be found when the considered processing operations are performed after the 3D reconstruction. In this case, the averaged standard scores of the gaussian smoothing operators applied over all voxels of the reconstructed 3D images are around 0 because smoothing is already performed in the FBP algorithm.

Finally, we compare the sensitivity of the quantitative analysis of the amyloid plaques in the mouse brain to DP compression and 8-bit conversion. Given the complete 3D prediction map, we compute the values of the standard scores for four plaque parameters: the total volume of plaque, the plaque load (which is the ratio of plaque volume to the total organ volume), the total plaque count and the plaque mean volume. The results, shown in Figure 4h, indicate that global segmentation parameters are conserved upon DP compression. In contrast, the 8-bit conversion provides around 5% deviation with respect to the raw values, confirming the alteration of the raw prediction also in this case.

In this work, we describe an experimental method capable of quantifying and confirming the effect of image compression on the outcomes of SL-assisted tasks in microscopy. Our approach relies on determining the SL predictive uncertainty from the intrinsic statistical noise of raw data. As this noise is unavoidable, in particular the shot noise, our approach sets a lower bound to the predictive uncertainty in SL-assisted decision-making processes.

We show that 16-to-8 bits conversion and 10:1 JPEG compression can alter SL outcomes by more than 10 predictive uncertainties. Interestingly, these distortions are more important in 3D applications: the use of 8-bit conversion brings to 5% and 15% prediction change in OPT and LS datasets segmentation, respectively, exceeding raw predictions by many predictive uncertainties. The different outcomes alterations in the two 3D cases are probably related to the different propagation of the raw data noise through the pre-processing pipeline. In contrast, we observe that alterations induced by the DP compression can be considered as statistically equivalent to those provided by the noise of raw images in both 2D and 3D cases. By respecting the raw pixel value statistics and providing compressed images with size reduced by a factor up to 10, this image format represents a valuable solution to computational and data management challenges associated to computer-vision automated tasks in microscopy.

Given the non-negligible effects of image compression in diagnostic applications, this work highlights the importance to archive raw unprocessed microscopy data, as well as trace image processing operations before use in SL models. Finally, our method can be generalized to any field where acquisition devices can be calibrated and raw data undergo processing before AI use.

# Figures

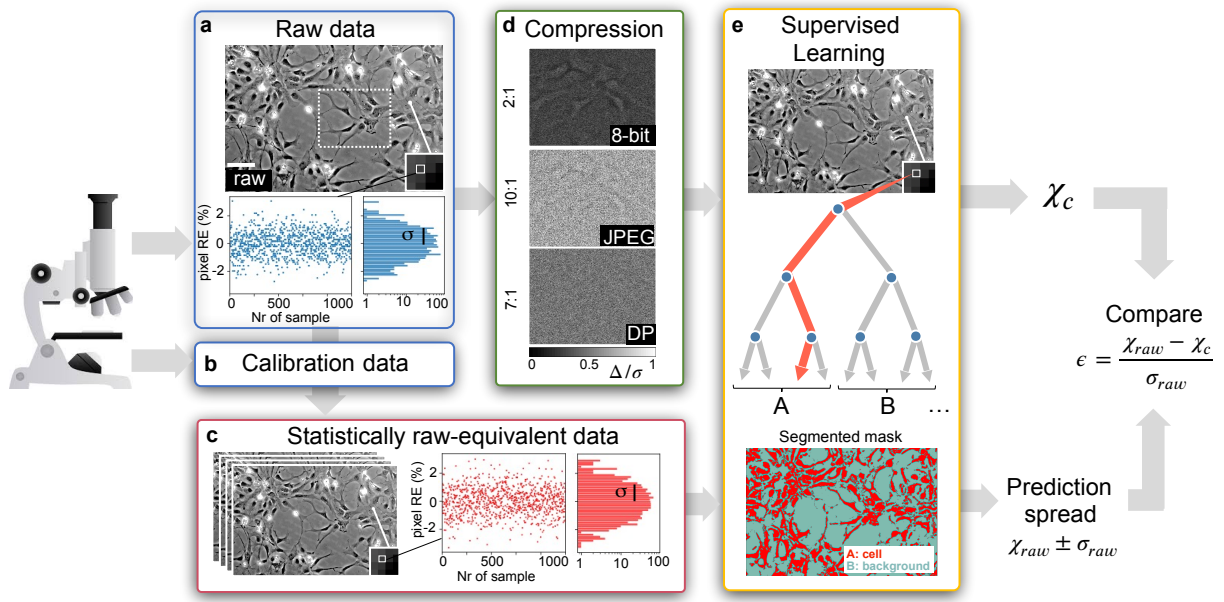

Figure 1 **Quantification of the effect of image compression on SL outcomes based on the predictive uncertainty estimation**

**a** Raw imaging data from a microscope are intrinsically affected by noise: pixel values in a raw micrograph (in the figure a PC image of human neural stem cells (scale bar: 100 μm)) have a statistical distribution of width σ and mean μ. This is shown by the plot of the relative error (RE) of the single pixel values obtained from 1000 acquired images at the corresponding light intensity (insets). **b, c** Via calibration data of the acquisition sensor, statistically raw-equivalent unprocessed images (with same σ and μ) are generated. **d** Image compression is performed on raw data: the differences between the 8-bit, 10:1 JPEG and 7:1 DP compressed formats and the raw image (Δ), normalized to the per-pixel noise σ (Δ/σ), indicate the artifacts induced by compression. The displayed image portions correspond to the dotted rectangle in **a**. **e** A pre-trained SL model (based for instance on a Random Forest algorithm), producing cell segmentation masks, is tested on the statistically raw-equivalent synthetic data to determine the standard deviation $\sigma_{raw}$ associated to a certain parameter $\chi_{raw}$. It also predicts a value of the considered parameter from the compressed data $\chi_c$. To verify the impact of image compression, the value of $\chi_c$ is compared to $\chi_{raw}$ via the standard score $\epsilon = \frac{\chi_{raw} - \chi_c}{\sigma_{raw}}$. If $|\epsilon| > 1$, the SL predictions on the compressed image exceed the statistical variability stemming from the noise of raw data.

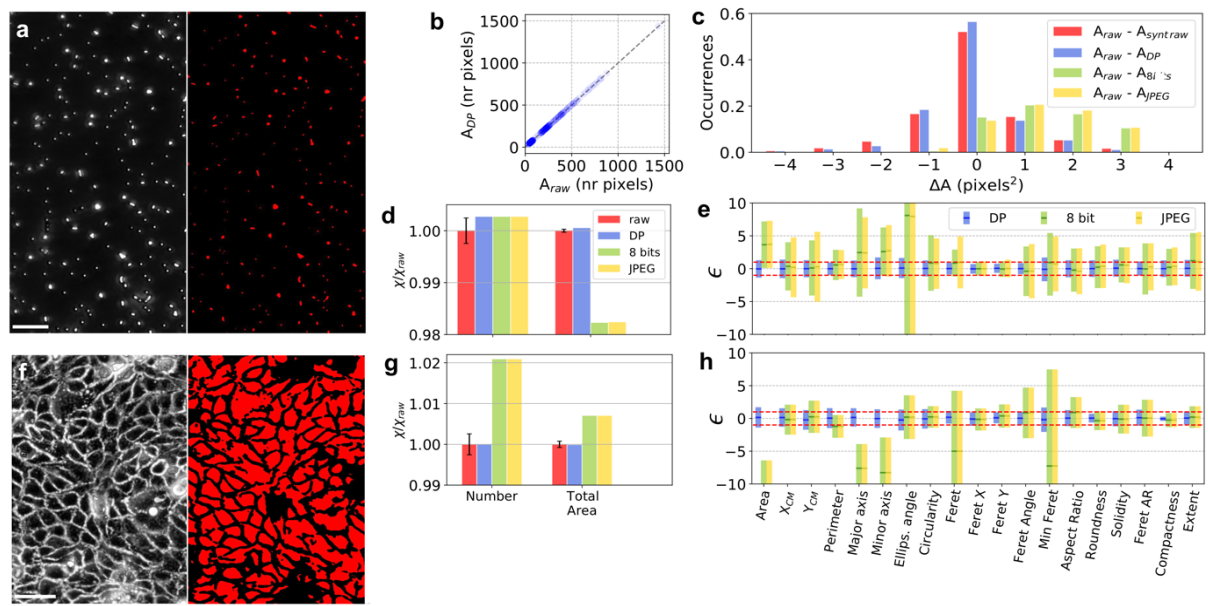

Figure 2 **Image compression tolerability in 2D segmentation SL tasks in microscopy**

**a** Micrograph of microspheres (scale bar: 20 μm) with corresponding segmentation mask. **b** Area of each segmented object obtained from the raw ($A_{raw}$) and the corresponding 6.7:1 DP compressed image ($A_{DP}$) for all objects. **c** Histogram of the difference in the single object area (ΔA) calculated from the raw and the statistically raw-equivalent images ($A_{raw} - A_{synt\,raw}$), the DP ($A_{raw} - A_{DP}$), the 8-bit ($A_{raw} - A_{8bit}$) and the 10:1 JPEG file ($A_{raw} - A_{JPEG}$). **d** Values of the parameters associated to the whole segmented image, such as the number of objects and total segmented area, obtained from the DP, the 8-bit and the JPEG segmented image and normalized to the raw value. The error bars on the raw values are obtained from the standard deviation of the values obtained from the synthetic raw images. **e** Average of the standard score ϵ of 19 parameters associated to the single segmented objects. **f** PC Micrograph of MPK cells (scale bar: 50 μm) with segmentation mask, downsampled to 8-bits and compressed into 6.1:1 DP and 10:1 JPEG formats. **g**, **h** Same as **d** and **e** for the MPK cells.

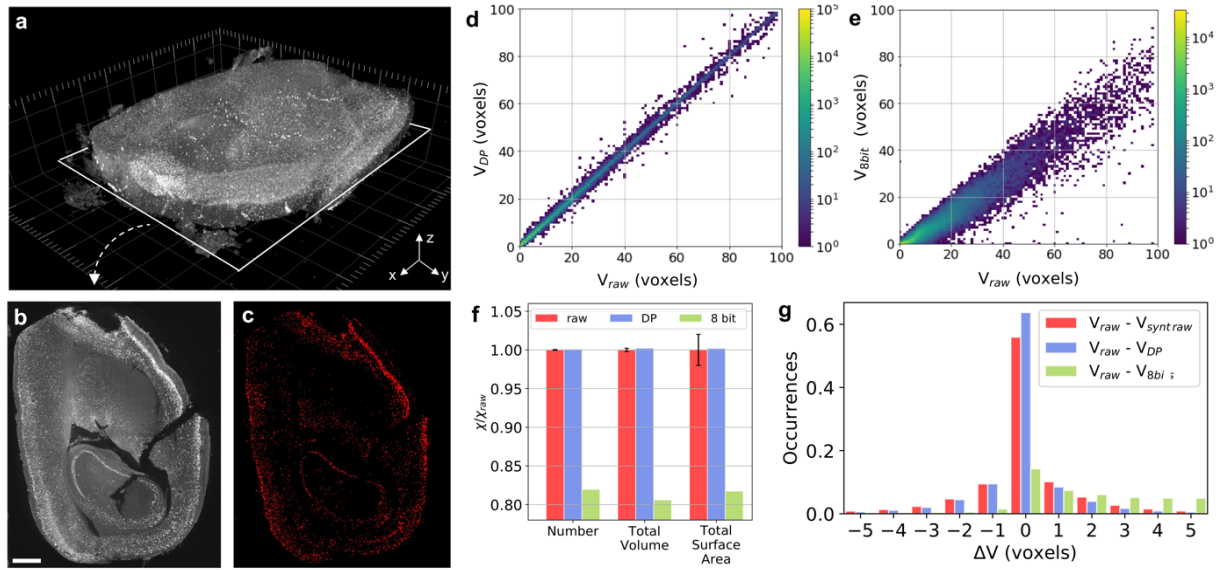

Figure 3 **Image compression tolerability in 3D segmentation SL tasks in light-sheet microscopy**

**a** 3D light-sheet image composed of 550 slices from a section of a mouse brain obtained after *c-fos* staining performed to identify neuronal nuclei. **b** Single slice of the 3D image (scale bar: 1 mm) **c** Corresponding segmentation mask. **d, e** 2D histogram comparing the volume of the same objects in the raw ($V_{raw}$) and 7.3:1 DP 3D compressed image ($V_{DP}$) (d), as well 8-bit 3D image ($V_{8\,bits}$) (e). **f** Values of the parameters associated to the whole segmented image, such as the number of objects, total segmented volume and surface area, obtained from the DP and 8-bit segmented image, normalized to the raw value. The error bars on the raw values are obtained from the standard deviation of the values obtained from the synthetic 3D images. **g** Histogram of the difference in the single object volume (ΔV) calculated from the raw and the statistically raw-equivalent images ($V_{raw} - V_{synt\,raw}$), the DP ($V_{raw} - V_{DP}$), and the 8-bit 2D image ($V_{raw} - V_{8bit}$).

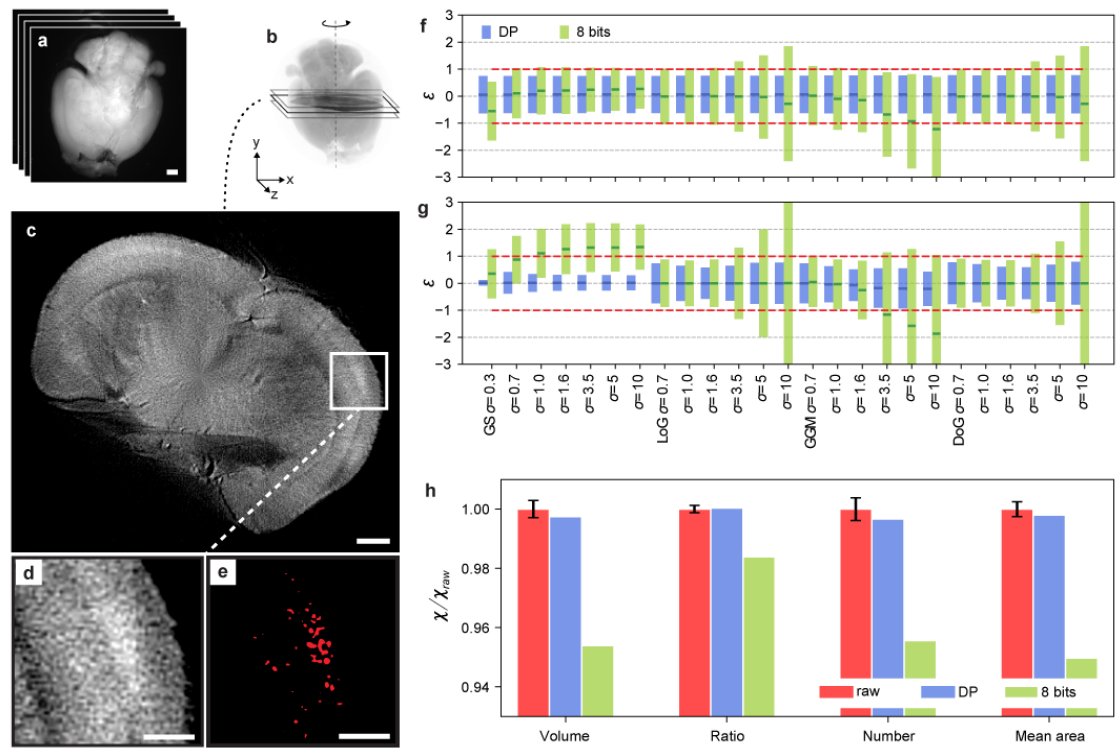

Figure 4 **Image compression tolerability in 3D segmentation of amyloid Alzheimer plaques imaged with optical projection tomography**

**a-c** Typical OPT pipeline. Fluorescent projections of an organ at different angles are used to reconstruct its 3D image via an inverse Radon transform. Transverse slices are used for physiological analysis (scale bar: 1 mm). **d** Close-up of a reconstructed slice showing amyloid plaque deposition in the mice brain. **e** Segmented mask of amyloid plaques (scale bar: 300 μm). **f, g** Plot of the single pixel standard scores of four image-processing operators for different sizes of the Gaussian-smoothing kernel calculated on projections (**f**) and reconstructed slices (**g**) obtained from the 7.8:1 DP compressed and the 8-bit converted 3D image. **h** Normalized values of amyloid plaque characteristics associated to the whole dataset, obtained from the raw (red), the DP compressed (blue) and the 8-bit converted datasets (green). The error bars on the raw values are determined via the synthetic statistically raw-equivalent 3D images.

**Acknowledgements**

The authors are grateful to F. Bugnon and C. Brack for technical support. This project has been partially funded by Innosuisse (grant no 31434.1 IP-ICT) and by the Horizon 2020 Framework Programme of the European Union with the project ADgut (grant no 686271).

**Authors contribution**

E. P. performed PC and LS segmentation analysis. C. S. performed OPT segmentation analysis. F. C. acquired PC images. D. N. and A. P. prepared samples and performed OPT measurements. A. T. prepared samples for LS microscopy. A. Ro. prepared samples for PC measurements. S. P. and L. B. provided LS datasets and contributed to LS microscopy data analysis. C.C. supported in the use of DP compression. T. L. co-designed the OPT system. T. L. and A. Ra. contributed to the idea of the paper and supervised the OPT data acquisition. B. S. contributed to the idea of the paper and in writing the article. E. P. and J. E. conceived the presented idea. J. E. supervised the project. E. P., C. S. and J. E. wrote the article.

**Methods**

**PC**

For PC microscopy measurements, we used two inverted PC microscopes (Axiovert 40C and Axiovert 25, Carl Zeiss Jena GmbH) equipped with 5x, 10x and 20x objectives and a calibrated CMOS camera (CM3-U3-31S4M-CS, Sony). Samples of carboxylate microspheres of 500 nm diameter (Polysciences) were obtained by deposing 5 μL of an aqueous solution with a concentration of $3.6 \cdot 10^8$ particles/mL on a 170 μm thick glass slide. After solvent evaporation, PMMA at 0.1g/mL was added to the sample, then centrifugated and dried. Microspheres were imaged with a 20x objective and used for segmentation tests, as well as for the measurement of the microscope's point spread function (PSF). Two cell lines with variable confluence have been cultured: human neural stem cells (HIP) (A3890101, ThermoFisher) and mouse kidney collecting duct cells (MPK)[29]. Both cell lines were cultured and grown on a cell treated plastic surface at a temperature of 37 °C with an air atmosphere enriched with 5% $CO_2$.

Micrographs of HIP and MPK cells were obtained with a 10x and 20x objective, respectively. For the measurement of the modulation transfer function (MTF), we imaged the 1951 USAF test target (R3L3S1P, Thorlabs Inc.) in bright field configuration with the 5x objective.

All segmentation tests on PC images were carried out via the trainable Weka segmentation ImageJ plug-in[30], using a RF algorithm that produces pixel-based segmentations via a classification performed through selected image features. To analyze the segmented masks, we used the "Extended Particle Analyzer" macro of the Biovoxxel toolbox in ImageJ[31], allowing to analyze the segmented objects according to a large variety of morphological parameters, shape descriptors and angle orientations. As the RF model was trained on a single 16-bit raw image, the synthetic images and the DP compressed were automatically suited to be tested by the model, while the 8-bit and JPEG files needed to be upsampled to 16-bit depth.

**LS**

For LS microscopy measurements, we used a home-built mesoscale single-plane illumination microscope[26]. This setup consists of a dual-sided excitation path using a fiber-coupled multiline laser combiner (405, 488, 561 and 647 nm, Toptica MLE) and a detection path comprising a 42 Olympus MVX-10 zoom microscope with a 1x objective (Olympus MVPLAPO 1x), a filter wheel (Ludl 96A350), and a calibrated scientific CMOS camera (Hamamatsu Orca Flash 4.0 V3). The excitation paths also contain galvo scanners for light-sheet generation and reduction of shadow artifacts due to absorption of the light-sheet. In addition, the beam waist is scanned using electrically tunable lenses (ETL, Optotune EL-16-40-5D-TC-L) synchronized with the rolling shutter of the camera. Sample was illuminated by one of the two acquisition paths. Image acquisition was done using custom software written in Python 3. Z-stacks were acquired at 5 μm spacing with a zoom set at 1.25X resulting in an in-plane pixel size of 5.26 μm (2048x2048 pixels). Excitation wavelength of the c-*fos* antibody was set at 647 nm with an emission filter LP 663 nm bandpass filter (BrightLine HC, AHF).

Concerning the sample preparation, mice were perfused with 4% PFA and tissue was post-fixed overnight in 4% PFA. Mouse brain were prepared for imaging following the iDISCO procedure described by Renier et al.[32]. To visualize a reporter of neuronal activity (c-*fos*), a c-*fos* antibody (synaptic

System Anti c-*Fos* CN226003) was used to label the brain 1:2000 (0.25 ug/ml). This was coupled to an anti-rabbit Alexa Fluor-647 (far-red spectrum) (5ug/ml). After clearing, brains were immersed in a 10 x 20 x 45 mm quartz cuvette filled with DiBenzyl Ether (RI 1.56).

For PSF measurements, fluorescent Tetraspeck microspheres 0.1 μm were diluted into 1% agarose. Excitation wavelength was set at 488 nm with an emission 530/40 nm bandpass filter (BrightLine HC, AHF).

The 3D segmentation tests on c-*fos* positive neuron nuclei were performed via a voxel classification workflow realized via the open-source image analysis software iLastik[33]. The training step consisted of a manual attribution of classes (anatomy, c-*fos* positive nuclei, background) to a few voxels of a reduced 3D volume. The trained algorithm could therefore predict the type of the remaining voxels in the full volume. To do so, the software attributed to each voxel a vector of features computed by typical image analysis operators and used RF as learning algorithm, where several decision trees were built by randomly picking computed features and searching for proper decision boundaries to separate between classes. The output was a 3D probability map where the value of each voxel corresponded to the likelihood to be a c-*fos* positive neuronal nucleus. The final nuclei segmentation was then realized by applying a threshold value of 0.5 to the 3D probability map. As done for the 2D segmentation tests, the RF model was trained on a single 16-bit raw image stack, while the 8-bit stacks are upsampled to 16-bit depth.

**OPT**

For OPT measurements, we processed datasets of full intact mouse brains previously studied in [34] using an SL approach to quantify amyloidosis of an Alzheimer's disease mouse model. The epi-fluorescent image projections were acquired with a custom mesoscopic OPT setup consisting of a calibrated CMOS camera (ORCA-Flash 4.0 V2, Hamamatsu) coupled to a 300-mm achromat objective lens providing 0.5X magnification of the sample[35]. The sample was mounted on a motorized rotation stage allowing for projection acquisitions over 360 degrees by steps of 0.3 or 0.9 degrees in approximately five minutes. The sample fluorescent signal was excited by a 420-nm LED light source illuminating the whole organ. In this configuration, the OPT setup had an isotropic pixel-limited resolution of approximately 50 μm over the whole organ, due to the physical pixel size of the camera. Each set of projections (1024x1024x1200 matrix for 1200 projections) were previously saved as uncompressed 16-bit stacked .tif files. The 3D reconstruction of the sample was achieved by applying a filtered back-projection (FBP)[36] to the raw and compressed projection sets with the Matlab iRadon function. To do so, we follow the same procedure as previously described in [35]. After reconstruction, the 3D image volumes were cropped along the three dimensions in order to remove background contribution and conserve the brain signal only. The intensity of each voxels was normalized over the volume using Fiji[37] and its contrast enhancement tool before SL segmentation to accommodate for differences in dynamic range.

The plaque segmentation process relied on the same voxel classification workflow used for the segmentation of *c-fos* positive neuronal nuclei realized with the software iLastik[33] on LS datasets. In this case, the three classes manually labeled in the training step were plaque, anatomy, and background. 12 slices (2D) from the full mouse brain were manually annotated (brain tissues and amyloid plaques), half of them were used to train the RF algorithm, while the other half served as a test set. The final plaque segmentation and quantification was then realized by applying a threshold value of 0.5 to the 3D plaque probability map. This pipeline was the same as that provided by the authors of a previous study[34].

## SUPPLEMENTARY MATERIAL

**Optical calibration procedure of a microscope camera**

The pixel value recorded in a microscope image depends both on the signal (mean number of photons impinging on the pixel) and noise. In general, the information content deriving from the signal cannot be distinguished from the entropy due to noise. During camera calibration, we project a series of specific mean photons numbers on the bare sensor pixels, i.e. we inject a signal known in advance, so that the noise model of the camera may be accurately determined. Our procedure is adapted from that present in the EMVA1288 standard[38].

The inside of polytetrafluoroethylene (PTFE) integrating sphere is illuminated by a white LED stable to better than $1/10^5$ over a range of output powers from 0.1 mW to 200 mW (Figure S1). The light intensity is measured through a 1cm x 1cm NIST-traceable photodiode connected to a calibrated 7-digit voltmeter placed on the surface of the sphere. The sensor is placed on the axis of the main 5 cm sphere aperture, at a 1 m distance.

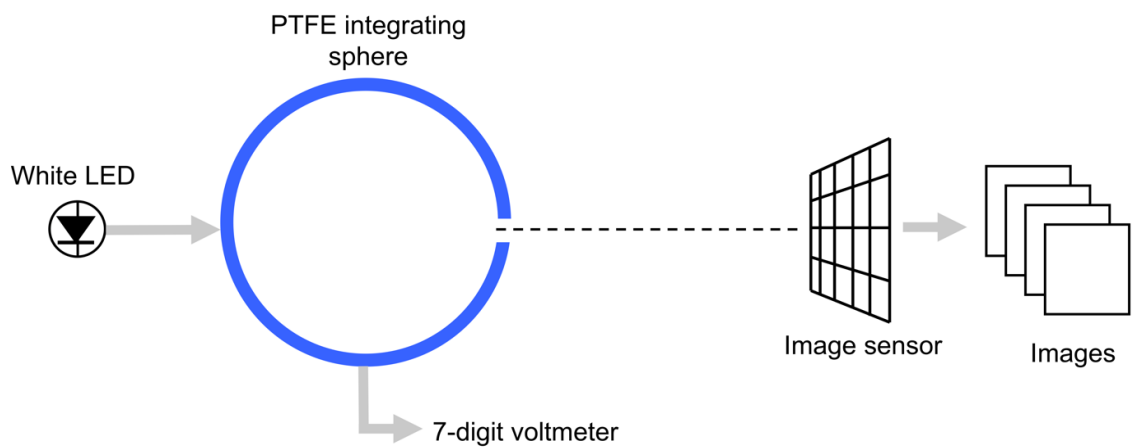

Figure S1 **Experimental setup for microscope cameras calibration**

A PTFE integrating sphere is illuminated by a white LED. A NIST-traceable photodiode is connected to calibrated 7-digit voltmeter. The microscope camera is placed on the axis of the sphere at 1 m from the 5 cm aperture of the PTFE sphere

In order to calibrate the camera sensor, 1000 images are acquired for each of 200 different illuminations, that are spaced according to a square-law from complete darkness to sensor saturation.

For each pixel we can therefore map each input light level (mean photon number $\langle n \rangle$) to a histogram of the recorded digital pixel values $d$, as shown in Figure S2a. From this plot a mathematical model of the sensor response can be formulated, as described in the EMVA1288. In particular, the relation between the standard deviation of the per-pixel noise $\sigma$ and the pixel value $d$ can be determined (Figure S2b).

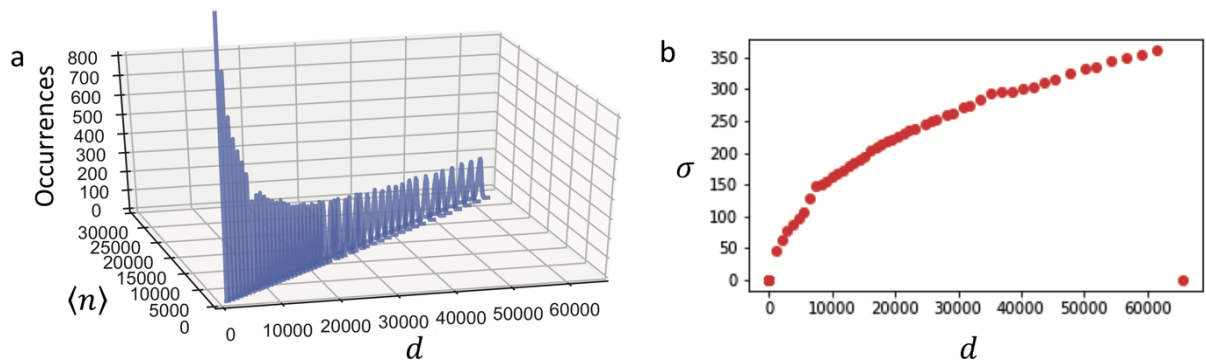

Figure S2 **Calibration curve of a single camera pixel**

**a** 3D representation of the statistical distribution of a pixel value $d$ for multiple input photon numbers $\langle n \rangle$. For a given $\langle n \rangle$, the number of occurrences is shown together with the digital pixel value $d$ returned by the sensor. **b** Plot of the standard deviation of the per-pixel noise $\sigma$ as a function of $d$ extracted from the statistical distribution shown in **a**.

**Resolution parameters in PC microscopy**

Resolutions parameters in a standard PC microscope, such as the Point Spread Function (PSF) and the Modulation Transfer Function (MTF), were measured to compare the predictive uncertainty of the tested AI segmentation models due to raw data noise with physical uncertainty quantities. We estimated resolution parameters from the raw and the DP compressed datasets, and show that their values are preserved upon DP compression.

We compare the value of the PSF of an optical microscope with 20x objective that is extracted from the raw and the DP images of 500 nm diameter microspheres. The FWHM of a single microsphere spatial profile, estimated over 10 different spheres, was measured as (882 ± 36) nm and (883 ± 36) nm from the raw and the DP images, respectively (Figure S3a), showing an excellent agreement between the results obtained with the two datasets. By deconvolving the measured profiles with a gaussian function of 500 nm FWHM, we obtain from both raw and DP compressed images a PSF with a 765 nm FWHM, corresponding to 4.5 pixels.

The MTF, which indicates the image contrast of a microscope,[39,40] was measured for an optical microscope with 5x objective in bright field with the help of a 1951 USAF target (Figure S3b). For each line series, the modulation is calculated as $M = \frac{I_{max} - I_{min}}{I_{max} + I_{min}}$, where $I_{max}$ (or $I_{min}$) is the maximum (or minimum) value of the averaged intensity profile of the lines (inset of Figure 2a). The value of M is calculated and averaged over 50 raw and corresponding DP images. The decrease of M with the spatial frequency obtained from raw and DP images is shown in Figure 2a, together with quadratic fitting lines. We obtain a cut-off frequency of $f_C$ = (285 ± 26) line pairs mm$^{-1}$ for the raw data and $f_C$ = (284 ± 25) line pairs mm$^{-1}$ for the compressed ones, confirming again an excellent agreement between the two datasets.

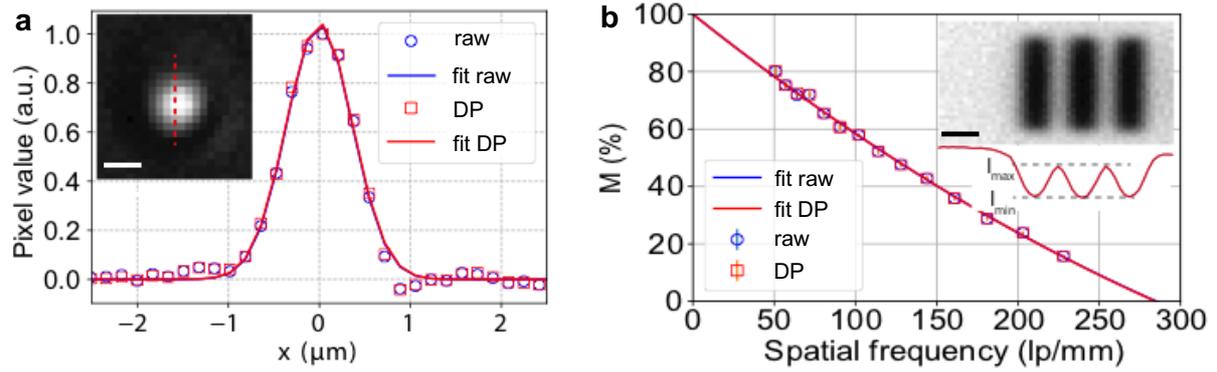

Figure S3 **Resolution parameters in PC microscopy**

**a** PSF obtained from the transversal profile of a polystyrene microsphere, imaged with a 20x objective (inset). Gaussian fits of the raw and DP data are used to estimate the FWHM. **b** Modulation transfer function estimated on 50 raw and on corresponding DP files as a function of line pairs mm$^{-1}$. The cut-off spatial frequency from raw and DP data is estimated with a polynomial fitting. Inset: a typical line series of a 1951 USAF target, from which the contrast of the averaged pixel intensity is measured.

**Resolution parameters in light-sheet microscopy**

We measure the PSF of the LS microscope from a 3D stack of 796 images of 100 nm diameter microspheres. In particular, we estimate the PSF from 10 microspheres selected in the raw dataset, as well as in the DP compressed one. As we are interested in comparison of the resolution parameters estimates from the raw and the DP compressed images, we did not implement the pre-processing pipeline utilized in [26] and do not exclude the possibility of beads aggregates.

The lateral PSF, obtained from the average of the FWHM of the x and y spatial profiles (Figure S4a and b), turned out to be (15.4 ± 5.1) μm and (15.3 ± 5.2) μm for the raw and the DP images, respectively. The axial PSF, i.e. the FWHM of the z profile (Figure 3e), is (20.1 ± 6.5) μm and (20.2 ± 6.7) μm for the raw and the DP images, respectively. We conclude that the values from the two datasets, as well as their statistical dispersion, are in very good agreement.

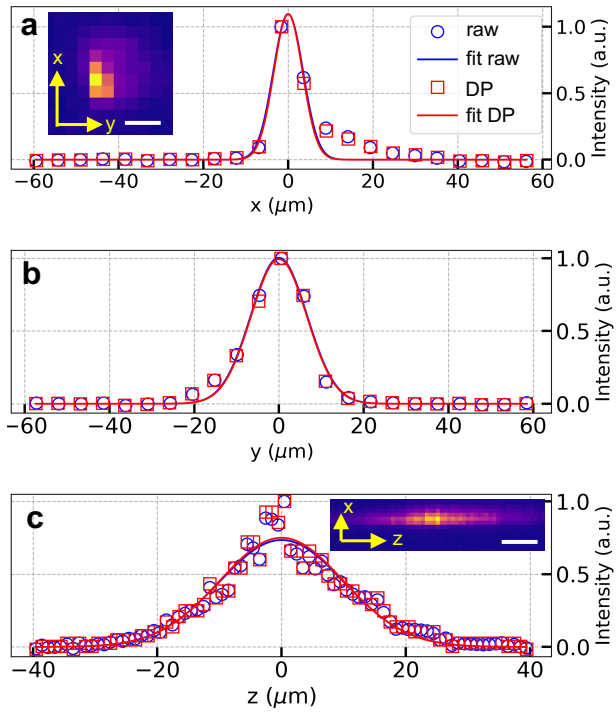

Figure S4 **Measurement of the PSF in light-sheet microscopy**

Lateral and axial PSF obtained from the spatial profile along x (a), y (b) and z direction (c) of a single 100 nm polystyrene microsphere imaged (insets) with a 1.25x objective.